# Superconducting Properties of 100 m Class Sr$_{0.6}$K$_{0.4}$Fe$_2$As$_2$ Tape and Pancake Coils

Xianping Zhang, Hidetoshi Oguro, Chao Yao, Chiheng Dong, Zhongtang Xu, Dongliang Wang, Satoshi Awaji, Kazuo Watanabe, Yanwei Ma

*Abstract*—Iron-pnictides are hotly studied since 2008 in the superconducting materials research area, due to their special properties and unclear mechanism. Big achievement has been made in the pnictide research during the past years. For practical uses, pnictide superconductor should be fabricated in a long wire form which can be used for different devices. In this work, 100 m class 7-core Sr$_{0.6}$K$_{0.4}$Fe$_2$As$_2$ (Sr122) tapes have been made using the powder-in-tube technique, which is reported for the first time. Clearly, an average $J_c$ of $1.3 \times 10^4$ A/cm$^2$ at 10 T was reached over the 115 meter length, showing a high property and good uniformity of 100 meter level Sr122 tapes. Using the 10 m long Sr122 tapes, two double-pancake coils were fabricated by a wind and reaction technique. No transport current could be measured for the coil made from the mono-filamentary tape. However, transport $I_c$ was obtained in the coil made from the 7-filamentary tape. The factors which affect the superconducting property of the coil were discussed in this work.

## I. INTRODUCTION

IRON-BASED superconductors [1] were hotly studied these days due to their special characteristics, such as the unconventional superconducting mechanism, high critical field, etc. The 122 type iron-pnictides, which was discovered in 2008 [2], are particularly attractive for the high field magnet applications, because of their superior transport critical current density $J_c$ higher than 1 MA/cm$^2$ achieved in thin films [3]-[5], high upper critical field $H_{c2}$ above 100 T with a small anisotropy <2, observed in single crystals [6]-[9]. For large-scale applications, superconducting wires with the ability to conduct high supercurrents are particularly desirable. In the past years, significant progresses towards high-performance iron-pnictide conductors were made [10]-[15], and the $J_c$ values of 122 wire and tapes have been rapidly improved to $10^5$ A/cm$^2$ at 4.2 K and 10 T [16], [17]. However, the high $J_c$ reported thus far have been limited to the short samples, which are just about 2 or 3 centimeters in length. For practical applications, long wires and demonstration devices should be fabricated. Generally speaking, compared to the short samples, there will be a $J_c$ deterioration in the long wire [18]. But the $J_c$ property of the long pnictide wire should be still acceptable for engineering uses. At the same time, whether the superconducting coils with good performance can be wind using the pnictide wire is another challenge.

In this work, we report the fabrication and superconducting properties of 100 m class Sr$_{0.6}$K$_{0.4}$Fe$_2$As$_2$ (Sr122) tapes. At the same time, coils were fabricated and tested. The test results are reported, including our understanding of the correlation between tape and coil performance.

Manuscript received September 6, 2016; accepted \*\*\*. Date of publication \*\*\*; date of current version \*\*\*. This work was supported by the National Natural Science Foundation of China (grant nos 51320105015, 51402292 and 51677179), the Beijing Training Project for the Leading Talents in S & T (grant no. Z151100000315001) and the Beijing Municipal Science and Technology Commission (grant no. Z141100004214002) *(Corresponding author: Yanwei Ma)*

Xianping Zhang, Chao Yao, Chiheng Dong, Zhongtang Xu, Dongliang Wang, and Yanwei Ma are with the Key Laboratory of Applied Superconductivity, Institute of Electrical Engineering, Chinese Academy of Sciences, Beijing 100190, People's Republic of China. (e-mail: ywma@mail.iee.ac.cn ).

Hidetoshi Oguro, Satoshi Awaji, and Kazuo Watanabe are with the High Field Laboratory for Superconducting Materials, Institute for Materials Research, Tohoku University, Sendai 980-8577, Japan.

## II. 7-CORE 100 METER LONG SR$_{0.6}$K$_{0.4}$FE$_2$AS$_2$ TAPE

### A. Experimental

The starting materials for synthesizing Sr$_{0.6}$K$_{0.4}$Fe$_2$As$_2$ precursor are Sr, K, Fe, As elements [19]. To compensate the loss of K element during the sintering process, additional K around 20 at.% was added. The prepared powders were put into the milling jar and mixed by high energy ball milling process. All procedures are handled in Ar atmosphere. After the ball milling, the powders were sealed in Nb tube, and then sintered at 900 °C for 35 h. The sintered bulk was ground into fine powder. To improve grain connectivity of the Sr$_{0.6}$K$_{0.4}$Fe$_2$As$_2$, 5 wt.% Sn was added into the powder. The final powder was filled in silver tubes, which were drawn into mono-filamentary wires and rolled to tapes ~0.35 mm in thickness [20]. For the 7-filamentary Sr122/Ag wires, the mono-filamentary wires were bundled into Ag tubes and finally deformed into 7-filamentary Sr122/Ag wires or tapes. The diameter of wires is about 1.5 mm, and the thickness of the tapes is around 0.4 mm. Finally, short samples were cut from the long tape, and then heated at 880 degree for 0.5 h. The transport current $I_c$ at 4.2 K and its magnetic field dependence were measured by the standard four-probe method with a criterion of 1 $\mu$V/cm.

### B. Results

Fig. 1 is one picture of the 100 meter long Sr$_{0.6}$K$_{0.4}$Fe$_2$As$_2$ tape. The tape has a good uniformity in both width and thickness. This indicates that the fabrication process is suitable for making the 100 meter class $Sr_{0.6}K_{0.4}Fe_2As_2$ tape.

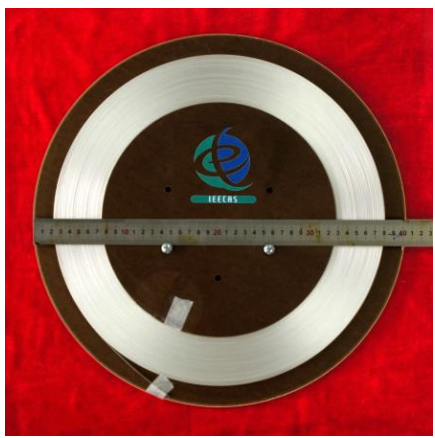

Fig. 1. Photograph of the 100 meter class Sr122 tape.

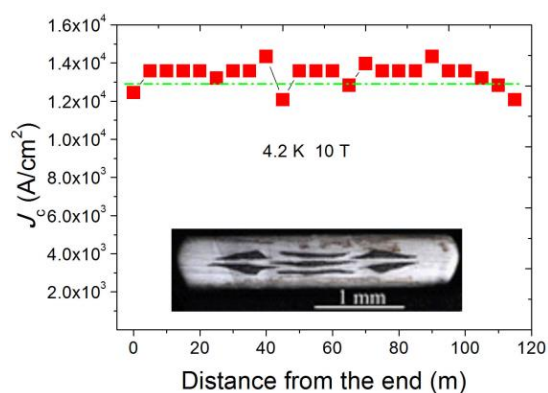

Fig. 2. Transport $J_c$ of the 115 m long 7- filamentary Sr122 tape. Inset: the cross-section image of the tape.

Fig. 2 exhibits the transport property of the 115 meter long $Sr_{0.6}K_{0.4}Fe_2As_2$ tape. Compared to the short samples made by the same procedure [19], the $J_c$ value is a little lower. For example, it is only about $1.3 \times 10^4$ A/cm$^2$ at 10 T and 4.2 K. But the uniformity is quite acceptable. The fluctuation of the $J_c$ was lower than 5%. This is good news for the coil construction, which usually requires a high homogeneity of the long wire.

## III. COILS MADE FROM SR122 SUPERCONDUCTOR

### A. Pancake coil made with a mono-filamentary tape

Specifications of the $Sr_{0.6}K_{0.4}Fe_2As_2$ tape and the double pancake coil are listed in Tables I and II. The pancake coil was wound using an Ag-sheathed $Sr_{0.6}K_{0.4}Fe_2As_2$ tape. The tape was insulated with mica tape, and the coil was solidified with epoxy resin at 80 degree. To get a good thermal conduction, the coil was covered with aluminum foil during the measurement.

TABLE I
PARAMETERS OF THE $Sr_{0.6}K_{0.4}Fe_2As_2$ TAPE

| Items | Parameters |
|---|---|
| Structure | Ag-sheathed Sr122 tape |
| Length | 10 m |
| Width | 5 mm |
| Thickness | 0.35 mm |
| Matrix | Ag |
| Number of filament | 1 |
| Insulation | Mica tape |
| Ic @ 4.2 K, 10 T | >100 A |

TABLE II
PARAMETERS OF THE COIL

| Items | Parameters |
|---|---|
| Structure | Double-pancake coil |
| Outside diameter | 118 mm |
| Inside diameter | 71.5 mm |
| Coil height | 12 mm |
| Turns number | 2×22 |

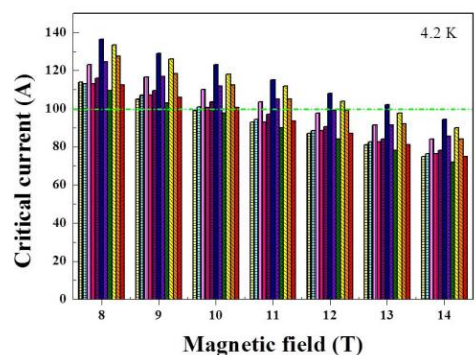

Fig. 3. Transport property of short samples evenly cut at 1 meter intervals from a 10 m long mono-filamentary tape.

Fig. 3 shows the $I_c$ value of the short samples evenly cut at 1 meter intervals of the 10 meter length tape. Clearly, the $I_c$ were over 100 A at 4.2 K, 10 T for all of the samples along the 10 meter scale. This means that the tape used for the coil fabrication have high uniformity and good superconducting property.

Fig. 4 shows one photograph of the coil after solidification by epoxy resin. Although the mica tape can sustain a high temperature, the color was changed to black after sintering. At the same time, mica tape is very brittle, and should be solidified with epoxy carefully. During the solidification process, vacuum pressure was applied.

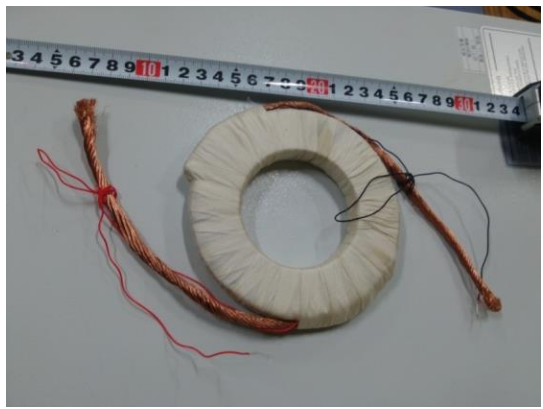

Fig. 4. Photograph of the coil after epoxy solidification. Outside of the coil was wrapped with silk tape for protection.

TABLE III
PARAMETERS OF THE $Sr_{0.6}K_{0.4}Fe_2As_2$ TAPE

| Items | Parameters |
|---|---|
| Structure | Ag-sheathed Sr122 tape |
| Length | 10 m |
| Width | 3.7 mm |
| Thickness | 0.44 mm |
| Matrix | Ag |
| Number of filaments | 7 |
| Insulation | Mica tape |
| Ic @ 4.2 K, 10 T | >10 A |

TABLE IV
PARAMETERS OF THE COIL

| Items | Parameters |
|---|---|
| Structure | Double-pancake coil |
| Outside diameter | 110 mm |
| Inside diameter | 71.5 mm |
| Coil height | 9.6 mm |
| Turns number | 2×15 |

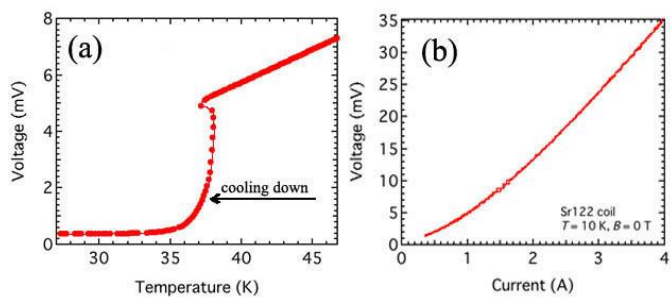

Fig. 5. (a) Temperature dependence of voltage of the coil made by mono-filamentary tape. (b) V–I characteristics of the coil made by mono-filamentary tape.

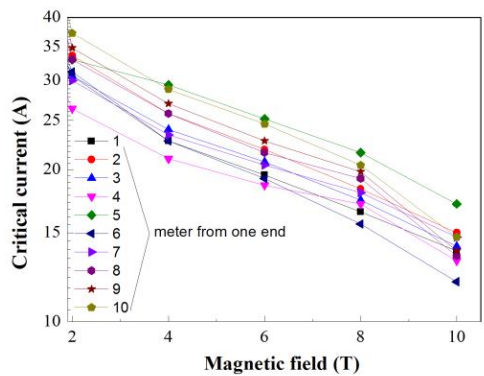

Fig. 6. $I_c$ performance of the short samples cut at 1 meter intervals from a 10 m long 7-filamentary tape used for coil winding.

By a GM cryocooler, the pancake coil was cooled down. During the temperature decrease, there is a voltage jump around 37 K, as exhibited in Fig. 5(a). This means that there is a supercurrent flow through the coil. But it should be noted that the coil never reach the zero resistance state in the measured temperature range. On the other hand, the nearly ohmic behavior indicates that there is no superconducting current flowing in the coil, as shown in Fig. 5(b). It is suspected that the superconducting core of the tape was broken, during either the mica tape wrapping or coil winding.

*B. Pancake coil made with a 7- filamentary tape*

To increase the mechanic property of the tape, we fabricated 7-filamentary tape for the coil winding. The specifications of the 7-filamentary $Sr_{0.6}K_{0.4}Fe_2As_2$ tape and the double-pancake coil are given in Tables III and IV.

Fig. 6 shows the transport property of the short samples evenly cut at 1 meter intervals from 10 m long 7-filamentary $Sr_{0.6}K_{0.4}Fe_2As_2$ tape. Although the $I_c$ currents have a fluctuation from end to end of the tape, they still keep a good superconducting property. The $I_c$ was over 10 A at 4.2 K, 10 T.

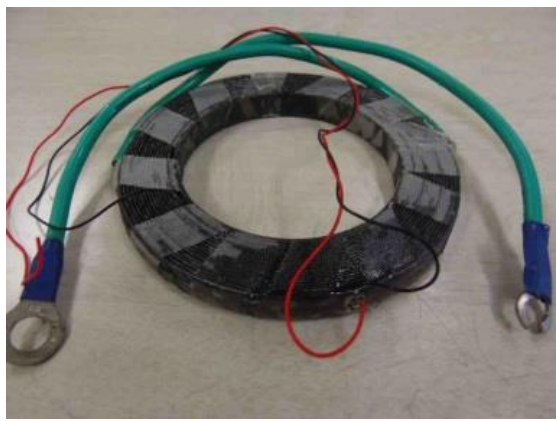

Fig. 7. Photograph of the coil after epoxy solidification. Part of the coil was wrapped with silk tape for protection.

Fig. 7 shows the coil solidified by epoxy resin. No redundant silk was used to get a good thermal conduction between the coil and the aluminum foil, which was put on the surface of the coil during the measurement.

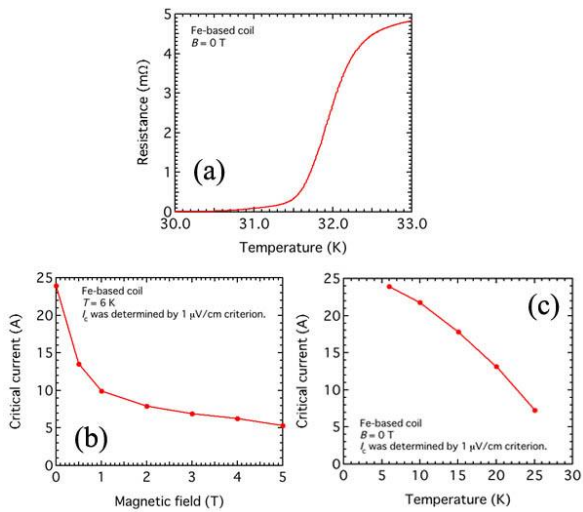

Fig. 8. (a) *R-T* curve of the coil made by 7-filamentary tape. (b) Transport property of the coil at different field. (c) Transport property of the coil at different temperature.

Under the conduction-cooled configuration, the coil was tested. The results are shown in Fig. 8. At about 33 K, there is a resistance decrease, and at around 31 K the transition was finished, as can be seen in Fig. 8(a). The $I_c$ is about 23 A at 6 K, 0 T, and decrease quickly at low field region. Then the $I_c$ was slowly decreased with increasing the field. This is similar to the behavior of the tape samples, which can hold a large critical current in high fields.

Compared to the short samples, the transport $I_c$ in the coil is relatively low. There are some possible reasons for the $I_c$ degradation in the coil:

(1) Bending diameter is not suitable for tape. Kovac has measured the strain property of the Sr122/Ag tapes, which have irreversible strains ~ 0.25% [21]. But it should be noted that the samples measured by Kovac and the tapes used for coil winding are different. Maybe the inner diameter of the coil is still smaller than required. There should have a strain measurement to the wire used for the coil winding.

(2) The strength of the Sr122/Ag tape is not strong enough for the automatically insulator wrapping and coil winding. During the insulator wrapping process, there is a violent vibration to the tape. This is very harmful for the brittle superconducting core. Wires with strong mechanical properties are needed to eliminate this negative factor and improve the property of the coil.

(3) Defect point in the long wire. Although we have got good results from the measurement of the short samples, it is difficult to say the whole long tape is good or not. It is needed to have an $I_c$ measurement for the long wire before it can be used for coil fabrication.

At the same time, the techniques applied for insulator wrapping, coil winding, solidification, etc., are also the possible reasons for the low superconducting property of the coils.

On the other hand, this is the first time that a coli is made using pnictide tapes. If we can solve the problems existed in the construction technique of the coil, the superconductivity of the coil will be enhanced.

## IV. Conclusion

We have successfully fabricated 100 meter class $Sr_{0.6}K_{0.4}Fe_2As_2$ tape for the first time. The average transport $J_c$ of the tape is about $1.3 \times 10^4$ A/cm$^2$ at 10 T and 4.2 K. At the same time, using 10 meter long $Sr_{0.6}K_{0.4}Fe_2As_2$ tapes, two coils were constructed by a wind and reaction method. For the coil made from 7-core tape, the transport $I_c$ was about 23 A at 6 K, 0 T. It is expected that the superconducting property of the coil will be improved after we can solve the problems existing in the construction technique of the coil.

## Acknowledgment

Xianping Zhang would like to thank Jingye Zhang, Dong Zhang, Tao Ma, Liwei Jing, Qianshan Ma, for their help in the winding and solidification of the coils.